\documentclass[accepted]{article}

\usepackage{microtype}
\usepackage{graphicx}
\usepackage{subfigure}
\usepackage{booktabs} 

\usepackage{hyperref}

\usepackage[accepted]{icml2022}

\usepackage{amsmath}
\usepackage{amssymb}
\usepackage{mathtools}
\usepackage{amsthm}
\usepackage[capitalize,noabbrev]{cleveref}

\usepackage{soul}

\theoremstyle{plain}

\theoremstyle{definition}

\theoremstyle{remark}

\usepackage[textsize=tiny]{todonotes}

\usepackage{titlesec}
\titlespacing*{\section}
{0pt}{0.45em plus 0.1em minus 0.1em}{0.08em plus 0.02em minus 0.02em}
\titlespacing*{\subsection}
{0pt}{0.3em plus 0.05em minus 0.05em}{0.04em plus 0.01em minus 0.01em}

\usepackage{xspace}
\def \ie{\textit{i.e.,}\xspace}
\def \eg{\textit{e.g.,}\xspace}

\icmltitlerunning{Emerging Patterns in the
Continuum Representation of Protein-Lipid Fingerprints}

\begin{document}

\twocolumn[
\icmltitle{Emerging Patterns in the
Continuum Representation of\\ Protein-Lipid Fingerprints}



\begin{icmlauthorlist}
\icmlauthor{~~Konstantia Georgouli}{pls}
\icmlauthor{~~Helgi I Ing\'olfsson}{pls}
\icmlauthor{~~Fikret Aydin}{pls}
\icmlauthor{~~Mark Heimann}{comp}
\icmlauthor{~~Felice C Lightstone}{pls}
\icmlauthor{~~Peer-Timo Bremer}{comp}
\icmlauthor{~~Harsh Bhatia}{comp}
\end{icmlauthorlist}

\icmlaffiliation{comp}{Center for Applied Scientific Computing, Lawrence Livermore National Laboratory,Livermore, California, 94550}

\icmlaffiliation{pls}{Physical and Life Sciences, Lawrence Livermore National Laboratory, Livermore, California, 94550}

\icmlcorrespondingauthor{Konstantia Georgouli}{georgouli1@llnl.gov}

\icmlkeywords{Machine Learning, ICML}

\vskip 0.3in
]



\printAffiliationsAndNotice{}  

\begin{abstract}
Capturing intricate biological phenomena often requires multiscale modeling where coarse and inexpensive models are developed using limited components of expensive and high-fidelity models. Here, we consider such a multiscale framework in the context of cancer biology and address the challenge of evaluating the descriptive capabilities of a continuum model developed using 1-dimensional statistics from a molecular dynamics model.
Using deep learning, we develop a highly predictive classification model that identifies complex and emergent behavior from the continuum model. With over 99.9\% accuracy demonstrated for two simulations, our approach confirms the existence of protein-specific ``lipid fingerprints'', \ie spatial rearrangements of lipids in response to proteins of interest. Through this demonstration, our model also provides external validation of the continuum model, affirms the value of such multiscale modeling, and can foster new insights through further analysis of these fingerprints.

\end{abstract}

\vspace{-1.5em}
\section{Intoduction}
\label{introduction}

Biological phenomena are intrinsically multiscale as many relevant interactions span through wide ranges of spatial and temporal scales~\cite{saunders2013coarse}. To capture such behavior at large- and long-enough scales while maintaining sufficient fidelity, multiscale modeling is used extensively~\cite{chopard2014framework}. Usually, multiscale techniques choose a few relevant scales and switch among them when and where needed~\cite{ingolfsson2022machine, bhatia2021machine}, thus offering tunable tradeoffs necessary to probe the complex dynamics of sophisticated systems.
Coarse-scale (approximate but fast) models are typically derived from finer-scale (high-fidelity but generally slow) models through simplification, aggregation, or changes in representation. Such approximations often lend themselves questions, not only about their accuracy but also about whether they are able to maintain and evolve key behavior of interest.
Here, we leverage deep learning (DL) to address such concerns by assessing the emerging behavior of a coarse-scale model.

Scientifically, we are interested in understanding the cancer initiation mechanism driven by RAS and RAF proteins, mutations in which are implicated in about a third of all human cancers and are responsible for over a million deaths each year. To study their signaling, multiscale model have been developed using (1) molecular dynamics (MD) simulations of these proteins and their interactions with the lipids on the plasma membrane, and (2) a continuum model that mimics these interactions by evolving densities of these lipids in response to the proteins~\cite{ingolfsson2022machine, stanton2021dynamic}.
Through a set of training MD simulations, lipid-protein interactions are represented as sets of 1-dimensional radial density functions (RDFs), which are then used to parameterize the continuum model. Given such a continuum model, the experts are interested in understanding how this model evolves the overall spatial dynamics of lipid-protein interactions, which include several more-complex considerations, such as 2-dimensional and nonisotropic interactions, multibody effects, discretization artifacts, and others.

We address this challenge using DL by casting the scientific question into a supervised learning task, essentially asking a neural network to learn if the emergent behavior from the continuum model is predictable with respect to the proteins. Representing the densities of different lipids as a multichannel image and the type and conformational state of proteins as ``class label'', we train a neural network to predict the label that corresponds to a given image. Using two different simulation sets (with six and twelve classes, respectively), we demonstrate exceptionally high predictive capability of our model. Scientifically, our approach deduces with high confidence that each type of protein (label) is associated with an easily distinguishable \textit{``lipid fingerprint''} (spatial arrangement of lipids around the proteins)~\cite{Corradi18}. Our results inform that this continuum model not only evolves different types of lipids with respect to the different types of proteins in ways that are not described by the input RDFs, but more importantly, is able to represent emergent yet predictable behaviors that can shed light into complex biological configurations of interest.

\section{Method}

\subsection{Description of the Continuum Data}
\label{Data}

Of interest to this work is the data generated through a macroscale, continuum model~\cite{stanton2021dynamic} that uses the dynamic density functional theory (DDFT)~\cite{marconi1999dynamic} to evolve densities of 14 species of lipids in response to two types of proteins --- RAS and RAF, which are represented as individual particles. The RAS protein can have two splice variants --- RAS4a and RAS4b --- each of which can exist in three membrane conformational states ($\alpha$, $\beta$, and $\beta'$). Whereas RAS proteins can exist by themselves, the RAF proteins exist on the membrane only when bonded with RAS to form a RAS-RAF complex. \autoref{sample-table} lists 12 possible combinations of these proteins, which we consider as class labels in this work.
Since the goal is to study lipid fingerprints~\cite{Corradi18} of proteins, we consider local, 30$\times$30 nm$^2$ neighborhoods of proteins, hereon referred to as ``patches''. By design, a RAS protein exists at the center of the patch, which provides a consistent frame of reference to explore fingerprints. We further restrict our study to patches that contain only one RAS protein or one RAS-RAF complex.

As described in \autoref{sample-table}, we consider two different simulations using similar continuum models. The first contains only the RAS4a splice variant and, hence, has only six class labels, whereas the second contains both RAS4a and RAS4b and, therefore, has 12 class labels. For brevity, we refer to these as P6 and P12 datasets, respectively. The lipid densities are sampled on regular grids: 37$\times$37$\times$14 (for P6) and 72$\times$72$\times$14 (for P12), with the last dimension corresponding to the 14 types of lipids in the system.
In order to account for the differences in the ranges of different lipid concentrations, the multichannel images were standardized per channel to zero mean and unit standard deviation.

\setlength{\tabcolsep}{3pt}
\begin{table}[!t]
\vspace{-0.65em}
\caption{Summary of the data distribution with respect to class labels for the P6 and P12 datasets. The datasets differ in the types of proteins (class labels) as well as the resolution of lipid densities.}
\label{sample-table}
\begin{center}

\begin{small}
\begin{tabular}{llrr}
\toprule
class ID & proteins (RAS state) & P6 & P12\\
\midrule
1 & RAS4b ($\alpha$) & 18.14\% & 5.01\% \\
2 & RAS4b ($\beta$)  & 57.52\% & 17.19\% \\
3 & RAS4b ($\beta'$) & 10.78\% & 2.80\% \\
4 & RAS4b-RAF (ma)   & 6.68\%  & 11.88\% \\
5 & RAS4b-RAF (mb)   & 6.20\%  & 12.69\% \\
6 & RAS4b-RAF (za)   & 0.62\%  & 0.42\% \\
7 & RAS4a ($\alpha$) & 0 & 4.51\%\\
8 & RAS4a ($\beta$)  & 0 & 15.82\%\\
9 & RAS4a ($\beta'$) & 0 & 4.67\%\\
10 & RAS4a-RAF (ma)   & 0 & 12.36\%\\
11 & RAS4a-RAF (mb)   & 0 & 11.91\%\\
12 & RAS4a-RAF (za)   & 0 & 0.73\%\\
\midrule
& Total data: & 2,492,720 & 320,000\\
& Data resolution: & 37$\times$37$\times$14 & 72$\times$72$\times$14\\
\bottomrule
\end{tabular}
\end{small}
\end{center}

\vspace{-2em}
\end{table}

\subsection{Description of the Deep Learning Model}

To demonstrate that the continuum model captures emergent behavior in the form of lipid fingerprints, we develop a classification model that can predict the protein classes, given an observed fingerprint. Specifically, we are given pairs $(X,Y)$ where the multichannel images $X$ represent lipid densities, and labels $Y$ represent the corresponding protein class. Our goal is to learn the mapping $X \to Y$ and we use a deep neural network to achieve this goal.

Given that the goal is to understand spatial patterns in lipid distributions, convolutional neural networks (CNNs)~\cite{hubel1963shape,hubel1970period} are a good fit as they are able to capture correlations across space and create a hierarchy of image features. In our case, a depth-wise convolution is also a useful tool since there exist correlations across channels (lipids) because (1) lipids are broadly either attracted to or repelled from the proteins, though with different densities, and (2) the amount of lipids packed into a region of space is limited, so high densities of some lipids lead to lower densities of others. We therefore utilize both types of convolutions in our model design.

After a few convolutional layers, we use dense layers to further reduce the data nonlinearly to accommodate remaining variations in data and create a $k$-dimensional latent space. We choose $k$ to be equal to the number of classes ($k=6$ and $k=12$ for the two datasets). This latent space generates unscaled log probablilties (logits), which are then converted to (normalized) class probabilities using a \textit{softmax} operation. We also utilize batch normalization layers as well as the \emph{relu} activation, benefits of both of which are well established in the community. We used a categorical cross-entropy loss, which is standard for multiclass classification problems. Our final model architecture is given below.

{\small{\texttt{\textbf{X} $\to$ 
SeparableConv2D(filters=6, depth\_mult=6, kernel\_sz=1, strides=1, {relu}) $\to$
BatchNorm $\to$ 
Conv2D(filters=16, kernel\_sz=3, strides=2, {relu}) $\to$
BatchNorm $\to$ 
Conv2D(filters=16, kernel\_sz=3, strides=2, {relu}) $\to$
BatchNorm $\to$ 
Flatten $\to$ 
Dense(shape=4*k) $\to$ 
Dense(shape=k) $\to$ 
\textbf{z}.}}}

For the model implementation, the sequential TensorFlow API was used. The model was trained for until the loss converged (about 50 epochs) with a batch size of 128 using the Adam optimizer~\cite{kingma2014adam}. For the training process, 10\% of the data was randomly selected to validate the models. To accelerate the training, Horovod framework~\cite{horovod} was used to support distributed, data-parallel training using (40 and 20 GPUs for the P6 and P12 datasets, respectively). For each epoch, the average values for accuracy (total and per class), and the average prediction loss over the different GPUs were measured as performance metrics.

\begin{figure*}
\centering
\vspace{-0.25em}
\includegraphics[width=0.85\textwidth]{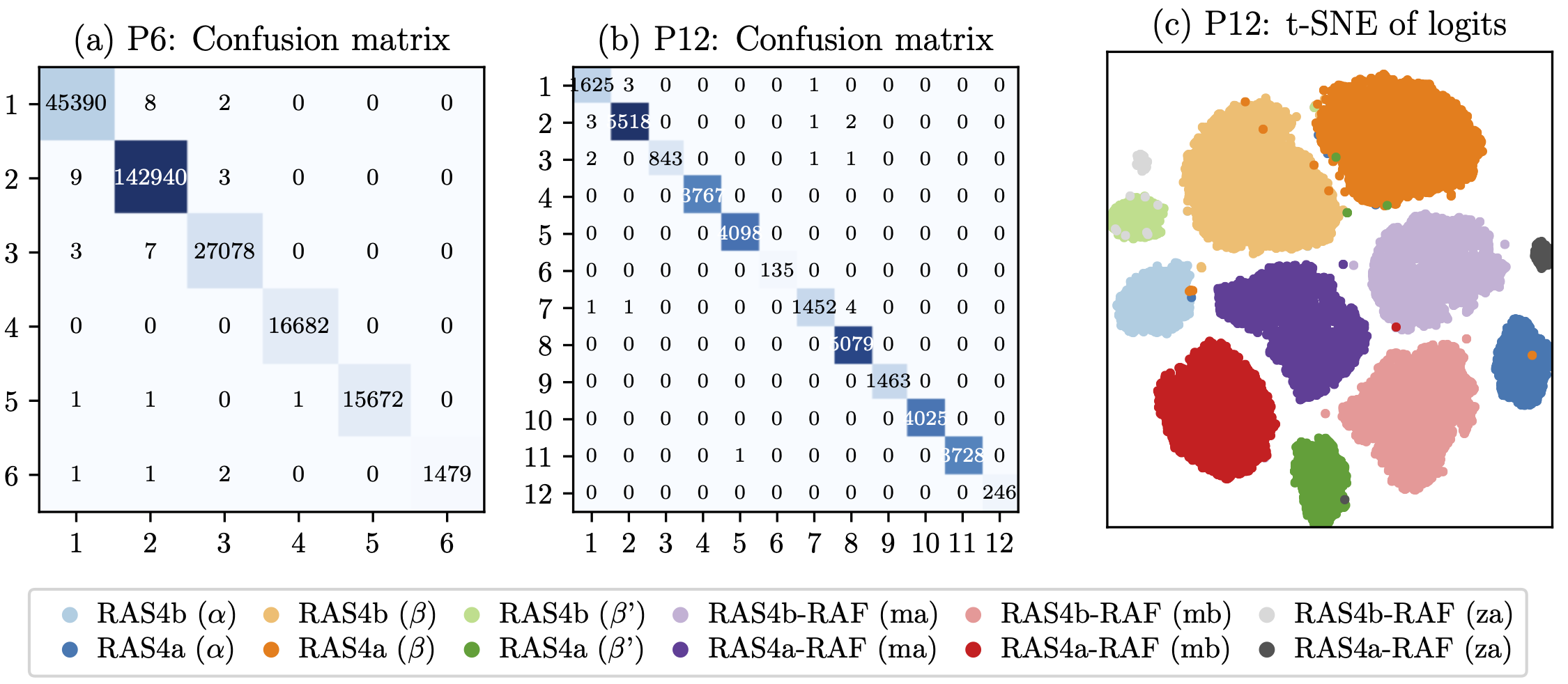}
\vspace{-1.em}
\caption{Our approach delivers high-quality prediction with over 99.8\% accuracy for both datasets, as highlighted by the corresponding confusion matrices. The figure also shows a 2-dimensional t-SNE visualization of 32,000 patches from the validation data and highlights the excellent separability of the different classes. Each point is a patch in the t-SNE space and colored by its predicted label.}
\label{fig:confmat}

\vspace{1.25em}
\includegraphics[width=\textwidth]{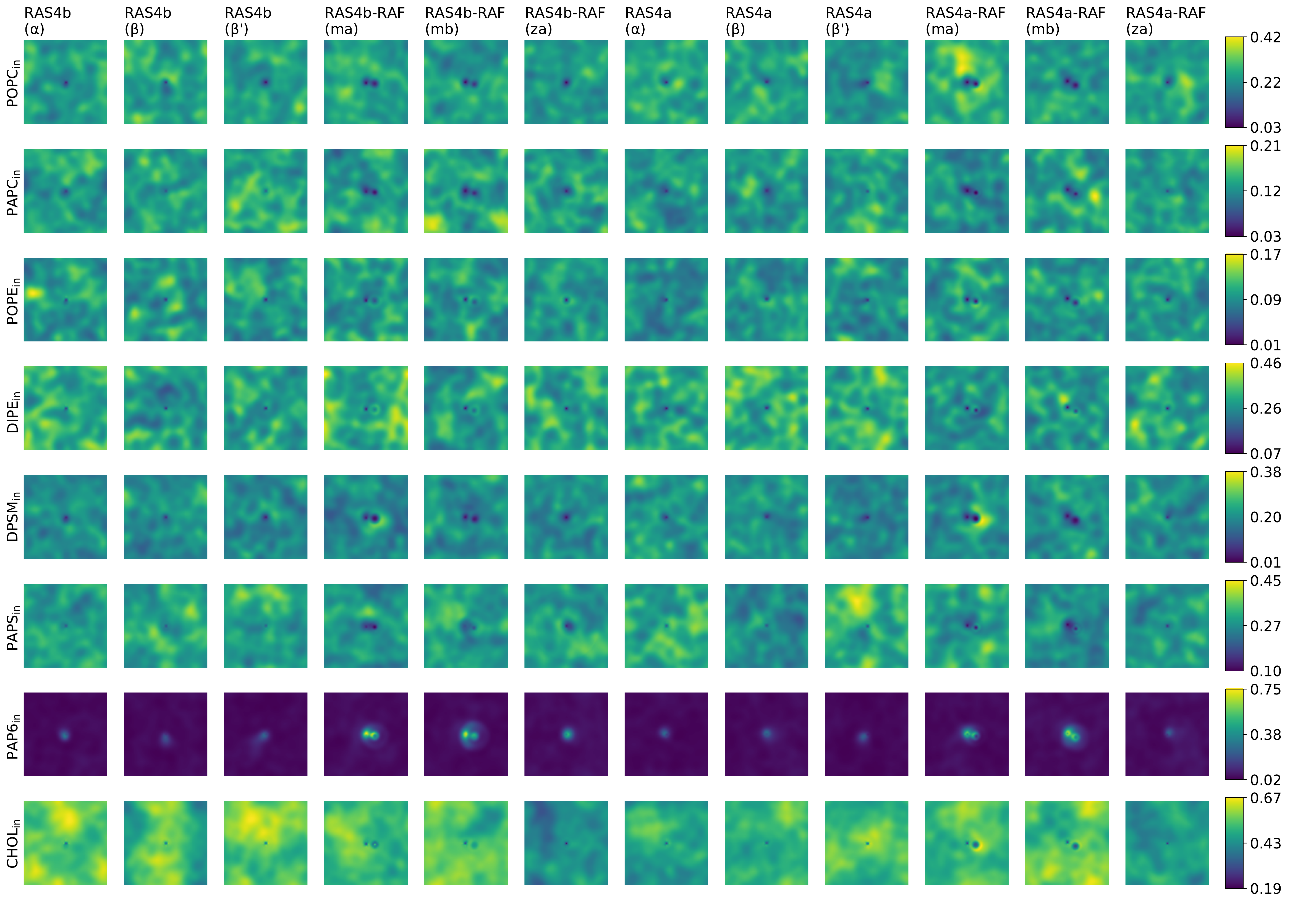}
\vspace{-2em}
\caption{We extract representative fingerprints (geometric median of clusters) for each protein state. The figure shows lipid fingerprints for P12: each row is a lipid in the patch and each column is a protein state. Colormaps are consistent across proteins for a given lipid.}
\label{patches_per_class12}
\end{figure*}

\section{Results}

\paragraph{P6: Classification model with 6 classes.}
We use our framework to predict the protein states from lipid \mbox{densities} (images) for the P6 dataset. Our model demonstrates exceptionally high predictive capabilities, as shown in \autoref{fig:confmat}(a). Specifically, we achieve a classification accuracy of 99.98\%, with only a handful of misclassifications, as indicated by the shown confusion matrix. We also note that our model is robust to this unbalanced dataset (see \autoref{sample-table}). Other commonly used metrics for classification, such as precision, recall, and the area under the precision-recall curve, also provide similar insights showing an excellent classification.

\vspace{-1em}
\paragraph{P12: Classification model with 12 classes.}
Next, we utilize our framework to study the larger, P12, system. Here, one of the goals was to introduce a new splice variant of RAS --- RAS4a --- and study how its fingerprints differ from RAS4b. We note that our model architecture generalizes well to this new dataset and the different input/output shapes (see \autoref{sample-table}). Simply changing the data sizes and using the same convolutional layers allowed us to train model for this dataset and achieve equally competent results. We note from \autoref{fig:confmat}, that even with on order of magnitude less data, we can still achieve over 99.9\% accurate classification.

For this experiment, we also visualize the clustering obtained through our model directly in the underlying space of logits (the model projects the input data onto a 12-dimensional space of logits, which are then transformed into class probabilities). Here, we compute a 2-dimensional t-SNE (t-distributed Stochastic Neighbor Embedding)~\cite{van2008visualizing} representation of this $k$ dimensional space and visualize all validation data as a scatter plot. We note from~\autoref{fig:confmat}(c) that our model separates the data neatly into 12 clusters --- each corresponding to a protein state, with only a handful of misclassifications. We note that many of the misclassified points in this visualization are likely due to the t-SNE projection error, and not as misclassifications of the model; for true count of misclassifications, one should refer to~\autoref{fig:confmat}(a).

Finally, toward the goal of capturing fingerprints, we study the canonical fingerprint of each protein class. Here, we visualize the geometric median of each cluster --- a patch that can be considered to be a representative of this cluster. \autoref{patches_per_class12} shows these representative patches and offers several interesting observations. Only the lipids on the inner bilayer are depicted since they show stronger responses than those in the outer bilayer (the inner bilayer of the membrane is directly in contact with the protein, as compared to the outer bilayer). First, as expected, different lipids show different responses to the proteins --- some show high density (yellow) near the proteins (center of the images) whereas others show low densities (blue). We also note that the local disturbance in lipids created by the proteins are different for each type of lipid, \eg POPC$_i$ shows two distinct ``blue dots'' whereas PAP6$_i$ shows a more aggregated ``halo'' (see rows RAS4b-RAF for states ma and mb). Finally, we also note that lipid PAP6 shows the strongest signal to noise ratio with a sharp increase in density near the protein. PAP6 also shows significant differences between proteins (RAS) and complexes (RAS-RAF).
Comparing RAS4a with RAS4b, we also note differences. Foremost, the fingerprints for RAS4a are a bit more diffused (compare PAP6 lipid). Similar pattern is identified in the comparison between RAS4a-RAF and RAS4b-RAF complexes.

\section{Conclusion}

In this work, we present a deep learning-based approach to learn protein-lipid fingerprints by predicting protein states given the lipids' spatial distributions. We show that our model generalized well to two simulation systems, with 6 and 12 types of protein states, including a new type of RAS splice variant. These simulations aim to understand the interaction behavior between lipids and RAS-RAF proteins. 
In both cases, our model exhibits exceptionally high classification accuracy --- more than 99.9\%. 
Our approach confirms that the given continuum model creates emergent behavior of lipids-protein interactions, given that proteins can be predicted based on the lipids. This model offers a way to establish the value of the underlying continuum model through external validation, and its potential in future work toward multiscale modeling. 
Based on the results of this preliminary work, a continuum description of lipids can be the basis for providing new insights for better understanding of complex biological mechanisms of interest such as the cancer initiation mechanism. 

In this work, we summarize the results of our prediction model for one protein and one complex systems. Going forward, we would like to expand this framework to incorporate more-complex systems with higher number of proteins.

\vspace{-0.8em}
\paragraph{Acknowledgements.}
This work has been supported by the Joint Design of Advanced Computing Solutions for Cancer (JDACS4C) program established by the US Department of Energy (DOE) and the National Cancer Institute (NCI) of the National Institutes of Health (NIH). This work was performed under the auspices of the US DOE by Lawrence Livermore National Laboratory under contract DE-AC52-07NA27344. Release number: LLNL-ABS-834976.


\clearpage
\bibliography{paper}
\bibliographystyle{icml2022}



\end{document}